%% file: main_prl_finalpub.tex
\pdfoutput=1
\documentclass[12pt,a4paper]{article}
\usepackage{overpic}
\usepackage[normalem]{ulem}
\usepackage{ifthen} 
\newboolean{pdflatex}
\setboolean{pdflatex}{true} 

\newboolean{articletitles}
\setboolean{articletitles}{true} 

\newboolean{uprightparticles}
\setboolean{uprightparticles}{false} 
\newcommand{\lhcborcid}[1]{\href{https://orcid.org/#1}{\hspace*{0.1em}\raisebox{-0.45ex}{\includegraphics[width=1em]{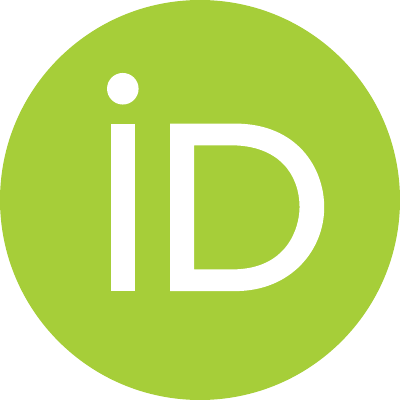}}}}

\def\paperauthors{LHCb collaboration} 
\def\paperasciititle{Evidence for B- -> D** tau nu decays} 
\def\papertitle{Evidence for $\Bm\to\Ddoublez\taum\neutb$ decays} 
\def\paperkeywords{{High Energy Physics}, {LHCb}} 
\def\papercopyright{\the\year\ CERN for the benefit of the LHCb collaboration} 
\def\paperlicence{CC BY 4.0 licence}
\def\paperlicenceurl{https://creativecommons.org/licenses/by/4.0/}

\newif\ifEnableSectionTOCLinks
\EnableSectionTOCLinksfalse 

\input{preamble}
\usepackage{longtable} 

\begin{document}

\renewcommand{\thefootnote}{\fnsymbol{footnote}}
\setcounter{footnote}{1}


\begin{titlepage}
\pagenumbering{roman}

\vspace*{-1.5cm}
\centerline{\large EUROPEAN ORGANIZATION FOR NUCLEAR RESEARCH (CERN)}
\vspace*{1.5cm}
\noindent
\begin{tabular*}{\linewidth}{lc@{\extracolsep{\fill}}r@{\extracolsep{0pt}}}
\ifthenelse{\boolean{pdflatex}}
{\vspace*{-1.5cm}\mbox{\!\!\!\includegraphics[width=.14\textwidth]{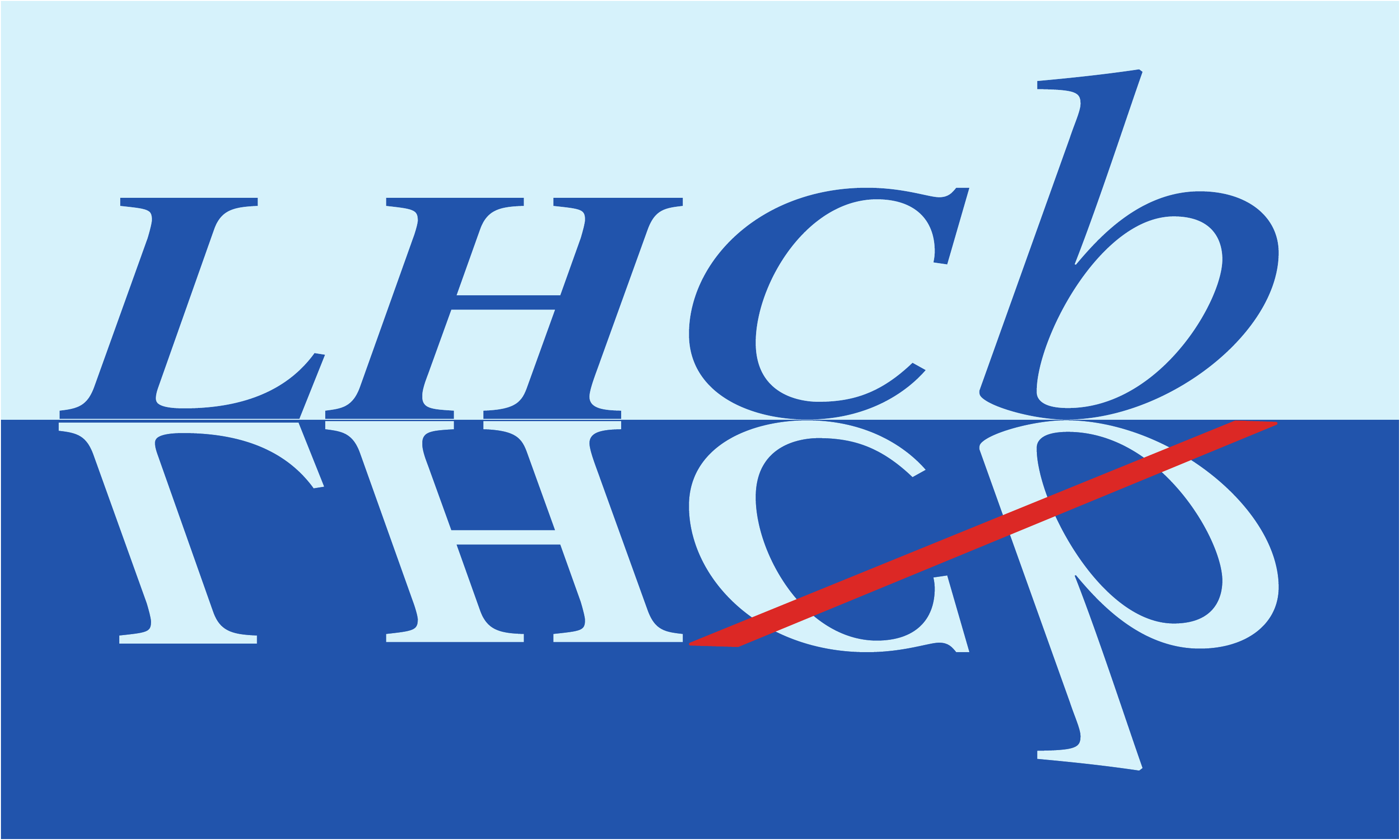}} & &}%
{\vspace*{-1.2cm}\mbox{\!\!\!\includegraphics[width=.12\textwidth]{figs/lhcb-logo.pdf}} & &}%
\\
 & & CERN-EP-2024-341 \\  
 & & LHCb-PAPER-2024-037 \\  
 & & 8 July 2025 \\ 
 & & \\
\end{tabular*}

\vspace*{3.0cm}

{\normalfont\bfseries\boldmath\huge
\begin{center}
  \papertitle 
\end{center}
}

\vspace*{1.0cm}

\begin{center}
\paperauthors\footnote{Authors are listed at the end of this paper.}
\end{center}

\vspace{\fill}

\begin{abstract}
  \noindent
The first evidence for the decay \mbox{$\Bm\to\Ddoublez\taum\neutb$} is obtained using proton-proton collision data collected by the LHCb experiment, corresponding to an integrated luminosity of $9 \invfb$, at centre-of-mass energies of 7, 8 and 13\tev. Here, the \Ddoublez meson represents any of the three excited charm mesons \donez, \dtwoz, and \doneprimz. The  \mbox{\decay{\Bm}{\Ddoublez\taum\neutb}} signal is measured with a significance of  $3.5\,\sigma$, including systematic uncertainties. 
 The combined branching fraction 
 \mbox{\BR($\Bm\to\dsumz\taum\neutb)\times\BR(\dsumz\to\Dstarp\pim)$}, where \dsumz denotes both \donez and \dtwoz contributions, is measured to be
\mbox{$(0.051\pm0.013\stat\pm 0.006\syst \pm 0.009(\rm{ext}) )\%$}, where the last uncertainty reflects that of the  branching fraction of the normalisation channel \mbox{\decay{\Bm}{\dsumz\Dsm^{(*)}}}. 
 The ratio  between the tauonic and muonic semileptonic \B decays, with the latter taken from world average values, is also determined and found to be \mbox{$ {\cal R}(\dsumz)=0.13\pm0.03\stat\pm0.01\syst\pm0.02\,(\rm{ext})$}.
\end{abstract}

\vspace*{1.0cm}

\begin{center}
  Published in Phys. Rev. Lett. 135 (2025) 021802.
\end{center}

\vspace{\fill}

{\footnotesize 
\centerline{\copyright~\papercopyright. \href{\paperlicenceurl}{\paperlicence}.}}
\vspace*{2mm}

\end{titlepage}


\newpage
\setcounter{page}{2}
\mbox{~}
%
%
%
%


\renewcommand{\thefootnote}{\arabic{footnote}}
\setcounter{footnote}{0}


\cleardoublepage


\pagestyle{plain} 
\setcounter{page}{1}
\pagenumbering{arabic}

\def\papertitle{
Evidence for $\Bm\to\Ddoublez\taum\neutb$ decays}
Lepton flavour universality (LFU) is one of the pillars of the Standard Model (SM) of particle physics. It postulates that the three charged leptons have exactly the same properties and couplings, except for their masses. However, LFU is  often violated in models beyond the SM~\cite{Fajfer_2012,Aebischer_2023}. Experimentally~\cite{Lees_2012,belle_collaboration_measurement_2015,belle-2016, the_belle_collaboration_measurement_2018,LHCb-PAPER-2017-017,LHCb-PAPER-2017-027, LHCb-PAPER-2022-052, LHCb-PAPER-2022-039, LHCb-PAPER-2024-007,belle-II}, evidence  of LFU violation is found in semileptonic $B$ decays, where the ratio of branching fractions ${\cal R}(D^{(*)})$, defined as  \mbox{$\BR(B\to D^{(*)}\taum\neutb)/\BR(B\to D^{(*)}\mun\neumb)$}, exceeds its SM prediction by 3.3 standard deviations ($\sigma$)~\cite{HFLAV}. This long-standing discrepancy  calls for more investigations of potential systematic effects. One of the largest systematic uncertainties common to all ${\cal R}(D^{(*)})$ measurements is the limited knowledge of the \Dstarp\taum\neutb contamination  by \mbox{$B\to\Ddouble\taum\neutb$} decays, where \Ddouble mesons decay to a state containing a  \Dstarp meson~\cite{HFLAV}. Hereafter, the \Ddouble symbol designates  all excited \cquark\uquarkbar and \cquark\dquarkbar mesons more massive than the $\Dstar(2010)$ state. The four lightest states are \ddzero, \done, \doneprim, and \dtwo with  masses 2343, 2422, 2412, and 2461\mevcc, respectively~\cite{PDG2024}. The last three mesons decay significantly into \Dstar\pion, while the first one decays only to $D$\pion. The \done and \dtwo states are narrow  with widths of 31 and 47\mevcc ~\cite{PDG2024}, while the other two are very wide with  widths in the range 200--300\mevcc. Theoretical predictions regarding the branching fractions of the \Ddoublez mesons to the \Dstarp\pim final state  can be found in Ref.\cite{rmp}. 

The goal of this  analysis is to search for  the  \mbox{$\Bm\to\Ddoublez\taum\neutb$} decays, and measure the branching fraction of the \mbox{$\Bm\to\dsumz\taum\neutb$} decay, where \dsumz represents the combined contributions from the \donez and \dtwoz mesons. The \dsumz mesons are reconstructed from the \Km\pim\pip\pip final state using the \mbox{$\dsumz\to\Dstarp\pim$}, \mbox{$\Dstarp\to\Dz\pip$}, \mbox{$\Dz\to\Km\pip$} decay chain.\footnote{The inclusion of charge-conjugate  processes is implied throughout, unless otherwise stated.} The \taum lepton is reconstructed using the \decay{\taum}{\pim\pip\pim(\piz)\neut} decay, where the \piz meson is not reconstructed.  The decay \mbox{$\decay{\Bm}{\dsumz\Dsm^{(*)}}$} is used as  normalisation channel for the branching fraction measurement. Its branching fraction is extracted from the recent analysis by the \lhcb collaboration~\cite{LHCb-PAPER-2024-001}. The normalisation channel is reconstructed using the \mbox{$\Dsm\to\pim\pip\pim$} decay that features the same vertex topology and  visible final state  of the signal decay. Many systematic uncertainties are in common between the two channels and  cancel  out in  the measurement of the ratio of branching fractions.
Using the known $\BR(B\to\dsumz\mun\neumb)$ value~\cite{PDG2024}, the ratio \mbox{${\cal R}(\dsumz)\equiv\BR(B\to\dsumz\taum\neutb$)/\BR($B\to \dsumz\mun\neumb)$} can also be determined and compared to the assumptions used in the  ${\cal R}(D^{(*)})$ measurements~\cite{Lees_2012,belle_collaboration_measurement_2015,belle-2016, the_belle_collaboration_measurement_2018,LHCb-PAPER-2017-017,LHCb-PAPER-2017-027, LHCb-PAPER-2022-052, LHCb-PAPER-2022-039, LHCb-PAPER-2024-007,belle-II}.

This analysis uses a data sample of proton-proton ($\proton\proton$)
collisions collected with the \lhcb detector, at centre-of-mass energies of 7, 8 and $13\tev$, corresponding to an integrated luminosity of $9\invfb$. The \lhcb detector is a 
single-arm forward spectrometer covering the pseudorapidity range 
$2 < \eta < 5$, 
described in detail in Refs.~\cite{Alves:2008zz,LHCb-DP-2014-002}.  The detector includes a high-precision tracking system
consisting of a silicon-strip vertex detector surrounding the $pp$
interaction region~\cite{LHCb-DP-2014-001}, and large-area silicon-strip detectors located upstream and downstream  of the $4{\mathrm{\,T\,m}}$ dipole magnet.
The minimum distance of a track to a primary $pp$ collision vertex (PV), the impact parameter (IP), 
is measured with a resolution of $(15+29/\pt)\mum$,
where \pt is the component of the momentum transverse to the beam direction, in\,\gevc. 
The online event selection is performed by a
trigger system~\cite{LHCb-DP-2012-004}, 
which consists of a hardware stage based on information from the calorimeter and muon
systems, followed by a software stage that performs a full event
reconstruction. Events are selected at the hardware stage if the particles forming
the signal candidate satisfy  a requirement on the energy deposited in the calorimeters or
if any other particle passes any trigger requirement. The software trigger requires a two-, three-, or four-track secondary
vertex with significant displacement from any PV and consistent with the decay of a \bquark hadron, or
a three-track vertex with a significant displacement from any PV
and consistent with the decay of a \Dstar meson. 
A multivariate algorithm~\cite{BBDT, LHCb-PROC-2015-018}  is used for the identification of secondary vertices. 
In the simulation, $pp$ collisions are generated using
\pythia8~\cite{Sjostrand:2006za,*Sjostrand:2007gs} 
with a specific \lhcb
configuration~\cite{LHCb-PROC-2010-056}.  Decays of hadronic particles
are described by \evtgen~\cite{Lange:2001uf}, in which final-state
radiation is generated using \photos~\cite{davidson2015photos}. 
The {\mbox{\textsc{Tauola}}\xspace} package~\cite{Davidson:2010rw} is used
to simulate the decays of the \taum lepton into \pim\pim\pip\neut and
\pim\pim\pip\piz\neut final states, according to the resonance chiral Lagrangian model~\cite{Nugent:2013hxa} with a tuning
based on the results from the \babar collaboration~\cite{Nugent:2013ij}. 
The interaction of the generated particles with the detector, and its response,
is implemented using the \geant
toolkit~\cite{Allison:2006ve, *Agostinelli:2002hh} as described in
Ref.~\cite{LHCb-PROC-2011-006}. 

This analysis follows the same  selection  regarding the reconstruction of \Dstarp and \taum candidates as the $ {\cal R(}\Dstar)$ measurement~\cite{LHCb-PAPER-2022-052}.
The presence of exactly one extra pion track, compatible with originating from the \Bm vertex is also required. Samples of right-sign (RS) or wrong-sign (WS) \Ddouble candidates are then formed by combining selected  \Dstarp candidates with a pion track compatible with originating from the \Bm vertex, having  opposite or same charge, respectively. The \Bm candidate is reconstructed from the combination of the \Ddoublez and \taum candidates. The pion and the \Dstarp decay products are required to form a good-quality vertex. 
The \Dz and \Dstarp candidates are required to have a mass close  to their known values~\cite{PDG2024}, and mass  sidebands are used to subtract the background. 
 As in previous analyses~\cite{LHCb-PAPER-2017-027, LHCb-PAPER-2022-052, LHCb-PAPER-2021-044}, a stringent requirement is applied to the displacement along the beam axis between the \taum and \Bm vertices, which must exceed four times the associated uncertainty. This selection is highly effective in suppressing the so-called ${\it prompt}$ background, due to \mbox{$B\to\Ddoublez\pim\pim\pip X$} combinations, where the pion triplet is produced at the $B$ decay vertex and $X$ denotes additional undetected particles (including zero if specified in parentheses).

 A  boosted decision tree~(BDT)~\cite{BBDT} classifier is trained to reject the background due to \mbox{$B\to\Dstarp\Dsm(X)$} events, where a \pim meson from a 5-prong \Dsm decay is combined with the \Dstarp to form the \Ddoublez candidate. The classifier exploits the different vertex topology between this background and the signal, and is  trained using simulated samples of the two  decays. This classifier is also efficient in reducing the prompt background, where the \pim meson used to form the \Ddoublez candidate and the pion triplet  are produced at the same vertex. 
 A second BDT classifier is used to reject remaining  fake \Ddouble candidates and is trained using simulated \mbox{$\Bm\to\donez\taum\neutb$} events as signal  and the WS data as background. 
 A further BDT classifier, referred to as \mbox{BDT-anti$D_s$}, is optimised to suppress the main background arising from \Dsm decays to three pions which mimics \taum decays. It uses the same inputs as for ${\cal R}(\Dstar)$ measurement \cite{LHCb-PAPER-2022-052}.  This classifier leverages the distinct decay dynamics of the signal \taum decays and the \Dsm decays, and is used in the signal extraction procedure.   

\begin{figure}[!tb]
\centering
\includegraphics[width=0.8\linewidth]{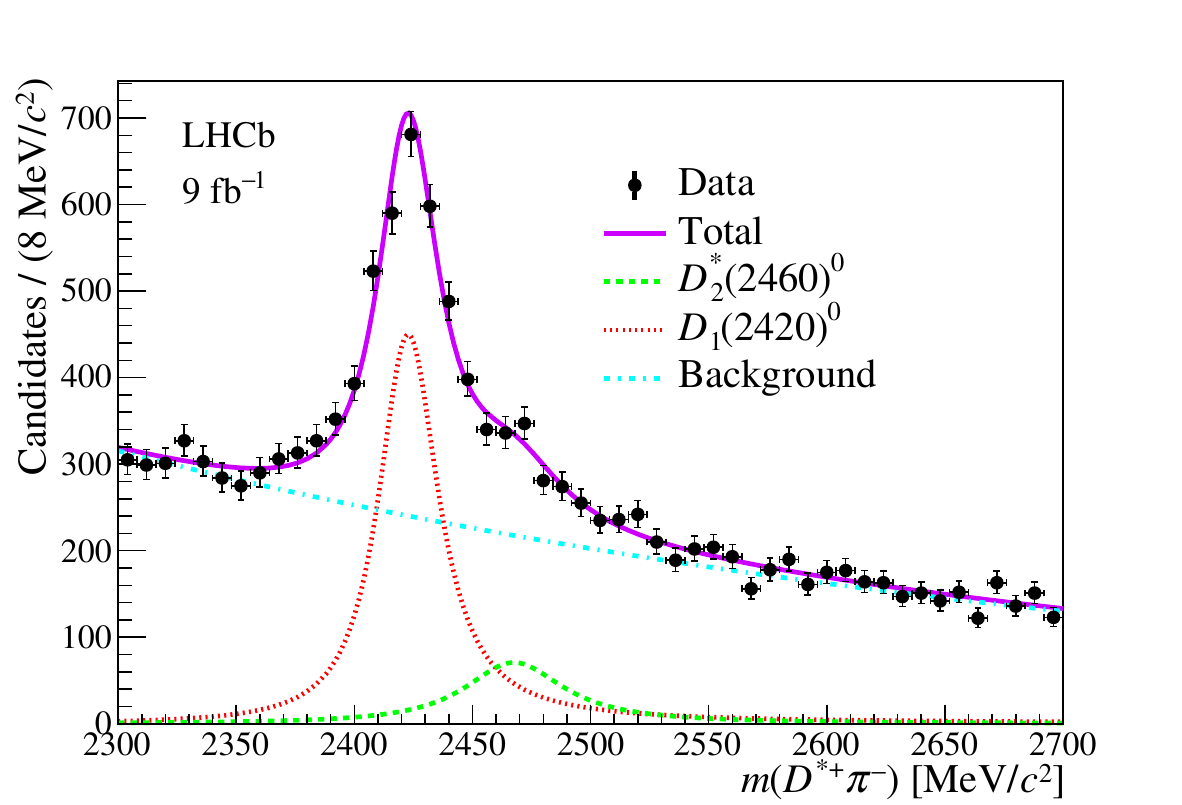}
\caption{Distribution of the \Dstarp\pim mass for all selected \Ddoublez\pim\pim\pip candidates in the full dataset, with the result of the fit also shown  A potential \doneprimz component can not be distinguished from the background.} 
\label{fig:doubledatafit_inclusive}
\end{figure}

 The \Ddoublez candidates mass distribution  after the final selection except the \mbox{BDT-anti$D_s$} requirement and the upper limit of 1600 \mevcc on the \pim\pip\pim mass   is shown in Fig.~\ref{fig:doubledatafit_inclusive}.
 To improve the resolution, the mass is calculated using the known \Dstarp mass~\cite{PDG2024}, $M_\Dstarp$, and the difference between the masses of the \Ddoublez and \Dstarp candidates, \mbox{$\dm \equiv m(\Dstarp\pim) - m(\Dstarp)$}, as $m(\Dstarp\pim)\equiv \dm + M_\Dstarp$. 
 An unbinned maximum-likelihood fit to the mass distribution is performed using two relativistic Breit--Wigner functions to describe the signal and one exponential function for the background. The yields of \donez and \dtwoz contributions are found to be $2456\pm75$ and $633\pm69$, respectively. These results  allow to determine the typical size and composition in terms of \donez to \dtwoz ratio of the initial \Ddoublez meson  sample, but are not explicitly used in the following.

\begin{table}[t]
\begin{center}
\caption{List of the various background categories. The first, second and third columns  indicate the decay chain producing each background,  the control sample dedicated to its study, and an estimate of its relative contribution in the selected sample, reflected by the number  of stars, respectively.}
\label{tab:backgroundlist}
 \begin{tabular}{l l  l}
    \hline
Background source&Control sample&Size\\ \hline
Fake \Ddoublez&&\\
~~Fake \Dz or fake \Dstarp&\Dz and \Dstarp sidebands&*\\
~~Genuine \Dstarp; fake \Ddoublez&WS sample&***\\
 &$B\to\Dstarp\Dstarm(X)$&*\\
&$B\to\Dstarp(DK)^{-}(X)$&**\\
\hline
Genuine \Ddoublez&&\\
~~$B\to\Ddoublez\pim\pim\pip X$ & $\Bm\to\Ddoublez\pim\pim\pip$ &*\\
~~$\Bm\to\Ddoublez\Dsm(X)$&$\Bm\to\Dstarp\Dsm(\to\pim\pim\pip)(X)$  &***\\
~~$\Bm\to\Ddoublez(DK)^{-}$&$B\to \Dstarp\Dz(\to\Km\pip\pip\pim)(X)$  &Negligible\\
\hline
 \end{tabular}
 \end{center}
\end{table}
Various background sources contribute to the \mbox{$\Bm\to\Ddoublez\taum\neutb$} selected sample. Their yields and distributions are determined using control data samples. Table~\ref{tab:backgroundlist} summarises the main background contributions and the corresponding control samples.
Fake \Dz or \Dstarp candidates are removed using  the sideband subtraction method.
Fake  \Ddoublez mesons can be formed combining a genuine \Dstarp meson with a random pion. 
This background source is  studied using the WS sample, leveraging the fact that random pions are equally likely to carry either charge.
Candidates from \mbox{$B\to \Dstarp\Dsm(X)$} and $\Dsm\to\pim\pim\pim\pip\pip$ decays and random \Dstarp\pim combinations fall into this category. 
Possible sources producing only WS combinations are investigated; the only significant one stems from the wrong reconstruction of the  slow pion from the \mbox{$\Dstarm\to\Dzb\pim$} decay into two almost identical tracks, one being used to form a \Dstarp candidate and the other to form a WS \Ddoublez candidate. Such cases are removed with a requirement on a dedicated multivariate classifier.
Additionally, fake \Ddoublez mesons can arise from \mbox{$B\to\Dstarp\Dstarm(X)$} decays, where the slow pion  is combined with the genuine \Dstarp meson to form the \Ddoublez candidate. They can also originate from \mbox{$B \to \Dstarp (D K )^- (X)$} decays, where the kaon is misidentified as a pion and is combined with the genuine \Dstarp meson. The latter two contributions are not included in the WS sample and must be added separately. These combined contributions represent a good description of the background as indicated from the  simulated sample  of inclusive $b$-hadron decays to final states containing   \mbox{$\Dstarp\pim\pip\pim$} that mimics the data as closely as possible as shown in Fig.~\ref{fig:deltaM_WS_check}.

\begin{figure}[!t]
 \centering
\includegraphics[width=0.8\linewidth]{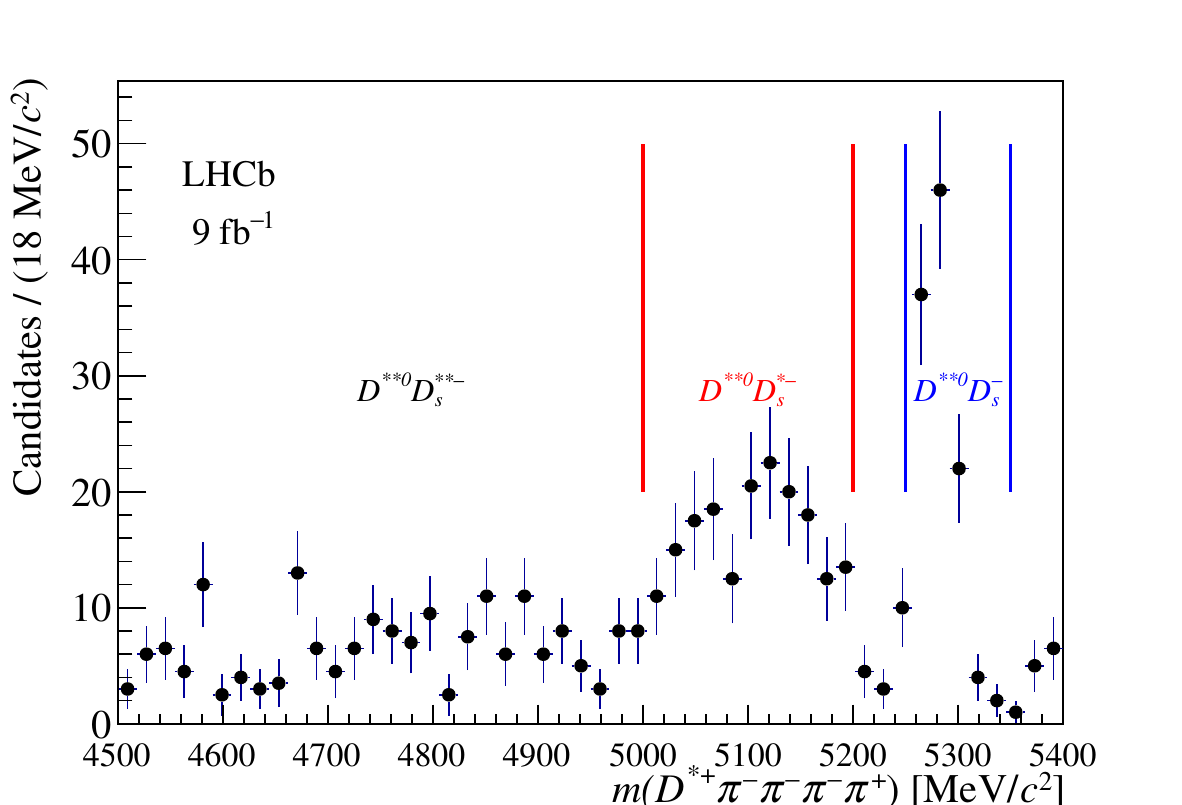}
\caption{ Distribution of the \Dstarp\pim\pim\pim\pip mass   for candidates with \pim\pip\pim mass consistent with the  \Dsm  mass. Background is subtracted using  candidates in the \Dsm mass sidebands. Three regions corresponding from left to right to \mbox{$\Bm\to\Ddoublez\Dsssm$}, \mbox{$\Bm\to\Ddoublez\Dssm$}, and \mbox{$\Bm\to\Ddoublez\Dsm$} are indicated by the vertical lines.}
\label{fig:doublestarDs}
  \end{figure}

The remaining background candidates consist of a genuine \Ddoublez candidate   associated with a fake \taum candidate. The largest source of this background is the {\it prompt} \mbox{$B \to \Ddoublez \pim\pim\pip X$} decay. However, its contribution is suppressed by a factor $10^{-3}$ due to the detachment requirement on the \Bm and \taum vertices, resulting in an estimated yield of $32\pm15$.
This background can be studied using a control sample obtained by removing the detachment selection requirement and analysing the exclusive decay \mbox{$B \to \Ddoublez \pim\pim\pip$}. 
The remaining  candidates arise mostly from  \mbox{$\Bm\to\Ddoublez\Dsm(X)$} decays, with \mbox{$\Dsm\to\pim\pip\pim(X)$} decays. The refined model used to describe the inclusive \Dsm meson decays to three and five pions is based on recent measurements perfomed by the \besiii collaboration~\cite{etapietappi,etaprho,etapipi0,etappipi0,eta3pi,kk3pi,omegapi,Ds3piX} and is identical to the one in the ${\cal R}(\Dstar)$ analysis ~\cite{LHCb-PAPER-2022-052}, as well as the weighting procedure described therein. The corresponding control sample is formed by selecting the exclusive decay  \mbox{$\Bm\to \Ddoublez\pim\pim\pip$} where the \pim\pip\pim mass is required to be within $\pm 30 \mevcc$ of  the known \Dsm mass~\cite{PDG2024}.
 A data-driven method is used to estimate the relative proportions of  \Dsm,  \Dssm and  \Dsssm mesons associated with a \Ddoublez candidate in  \Bm decays, which are an important input for a correct background description but are poorly known. The method uses a data sample obtained without the BDT-anti\Ds selection, but requiring the \Dsm fully reconstructed in the \pip\pim\pim final state.
Figure~\ref{fig:doublestarDs} shows the \Dstarp\pim\pim\pim\pip mass distribution  of the selected candidates. One can identify three regions from left to right: the \mbox{$\Bm\to\Ddoublez\Dsssm$}, \mbox{$\Bm\to\Ddoublez\Dssm$}, \mbox{$\Bm\to\Ddoublez\Dsm$} domains defined with $\Ddoublez3\pi$ mass in the ranges below 5000\mevcc,  \mbox{5000--5200\mevcc}, and \mbox{5250--5350\mevcc}, respectively. Together with the results obtained in  Ref.~\cite{LHCb-PAPER-2024-001} using the \mbox{$\Bm\to\Dstarp\Dsm\pim$} and \mbox{$\Dsm\to\Km\Kp\pim$} decay chains, this information is  used to constrain the yields of the various \Dsm contributions. 
Furthermore, the yields of \mbox{$\Bm\to\Ddoublez\Dssm$} and \mbox{$\Bm\to\Ddoublez\Dsm$} are determined to provide the normalisation for the branching fraction measurement. The fraction of \dsumz candidates in these yields are derived from the fit fraction of Ref.~\cite{LHCb-PAPER-2024-001}.

  Background due to \mbox{$\Bm\to\Ddoublez(DK)^{-}$} decays, where the notation $(DK)^{-}$ represents all combinations of a charmed meson  and a kaon  with a negative total charge, could also contribute, but this source  is expected to be highly suppressed due to the limited phase space available. 
Its contribution is studied using control samples obtained requiring a \Dstarp candidate and either a \mbox{$\Dzb\to\Kp\pim\pim\pip$}   or a \mbox{$\Dm\to \Kp\pim\pim$} candidate, where the \Kp track is consistent with originating from the \pim\pip\pim vertex. 
The first sample, consisting of $42\pm15$ \mbox{$\Bm\to\Ddoublez \Dzb K^{-}$} decays where the \Km track is not reconstructed, as can be seen in Fig.~\ref{fig:dz_deltam}, corresponds to an expected yield of $5\pm3$, given the relative efficiencies  between the control and signal sample~\cite{LHCb-PAPER-2022-052}. 
The second sample leads to a similar contribution. As a result, the background including a genuine \Ddoublez is assumed to consist only of the  {\it prompt} \mbox{$B \to \Ddoublez\pim\pim\pim X$} and the \mbox{$\Bm \to \Ddoublez\Dsm (X)$} contributions. 

\begin{figure}
\includegraphics[width=0.48\textwidth]{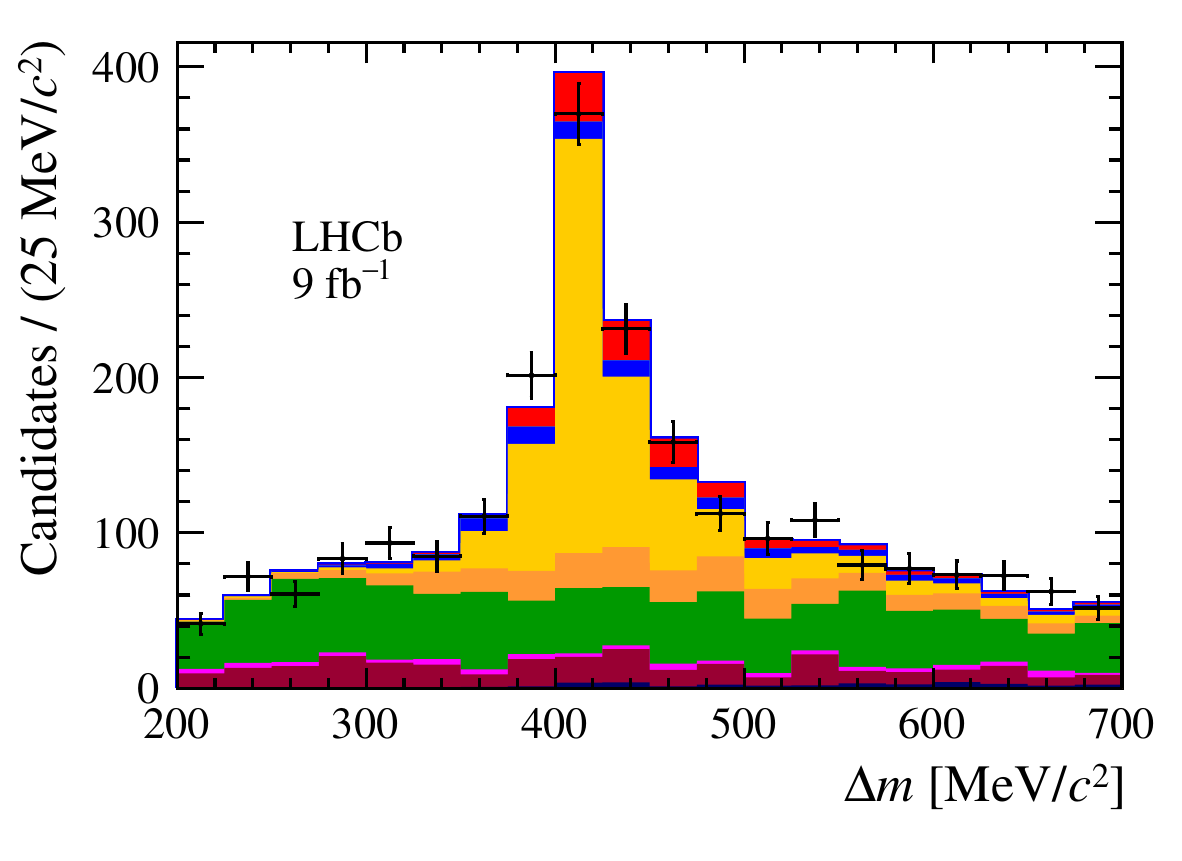}
\includegraphics[width=0.48\linewidth]{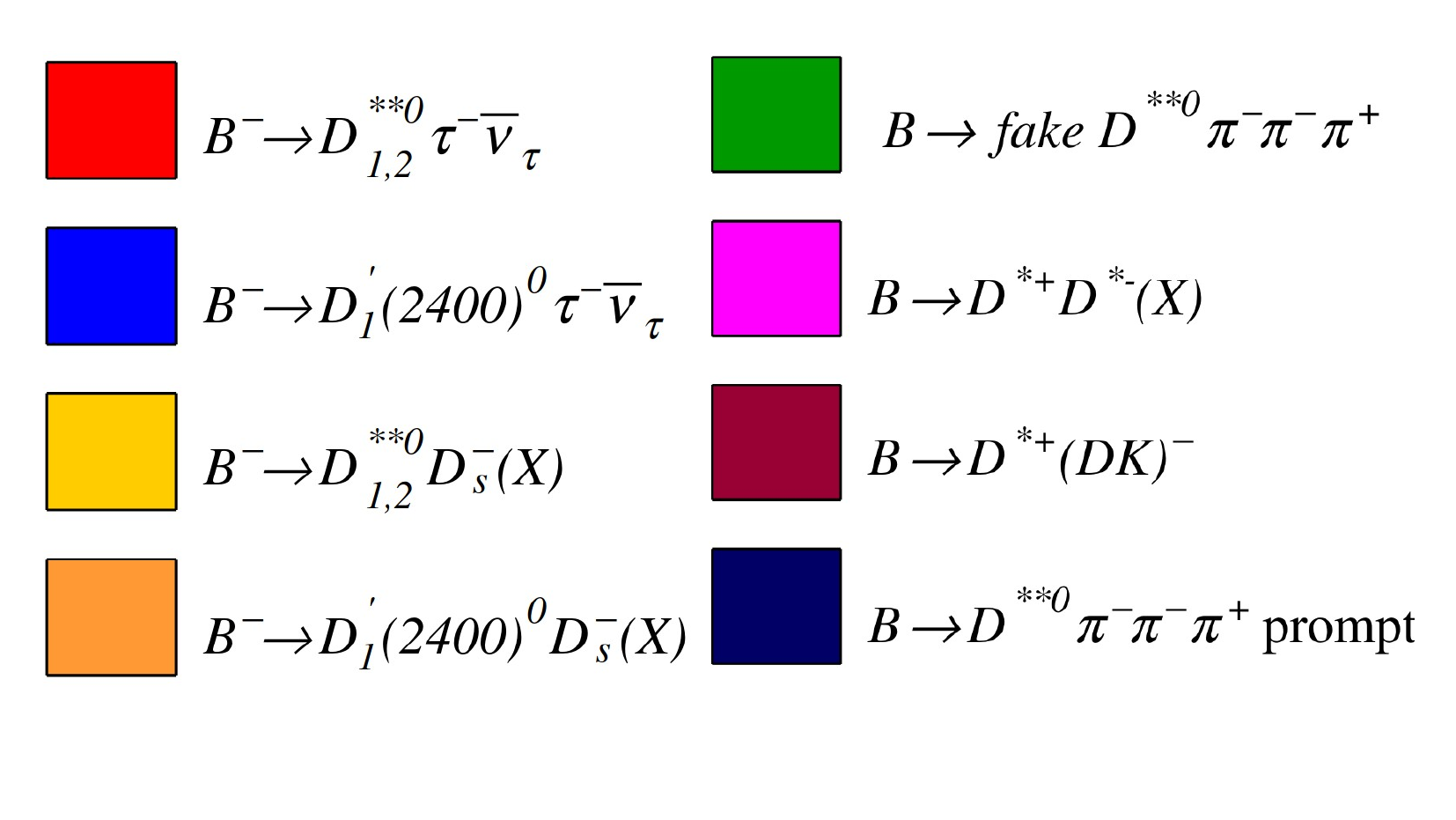}
\includegraphics[width=0.48\linewidth]{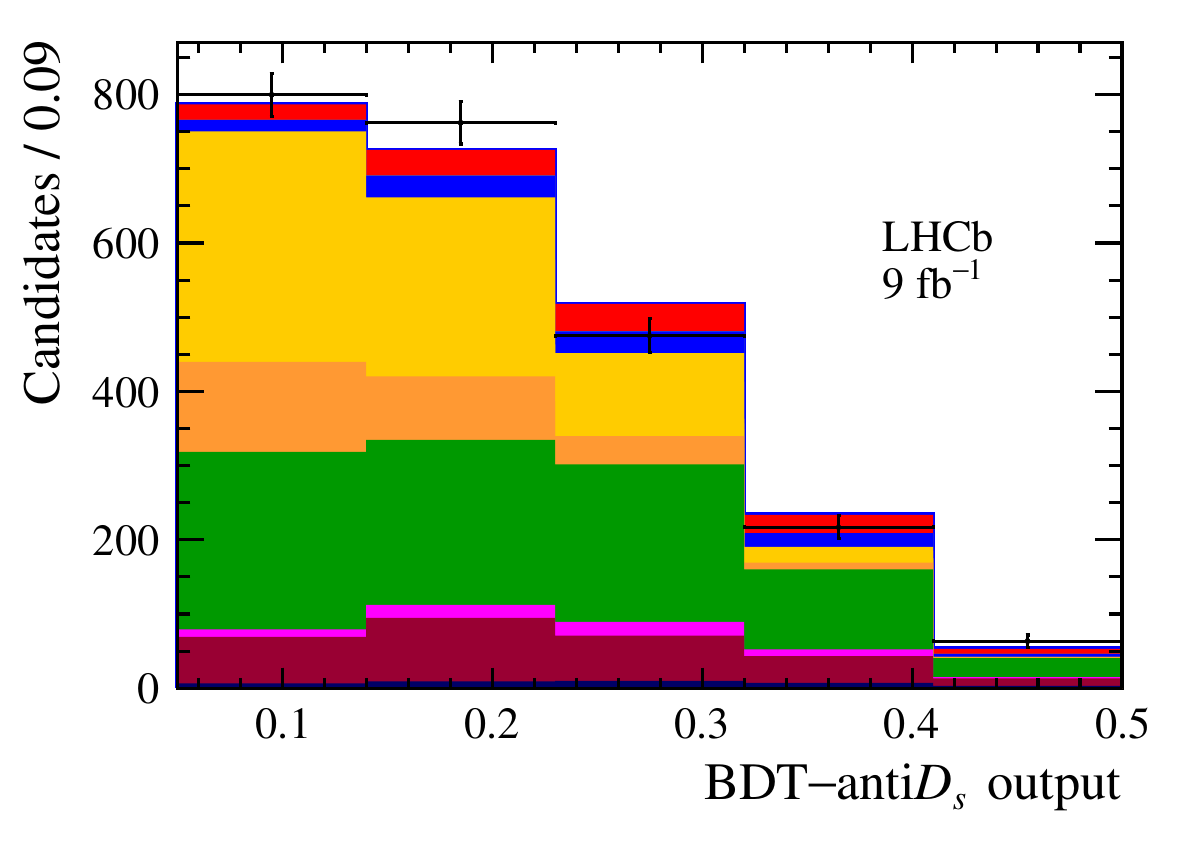}
\includegraphics[width=0.48\linewidth]{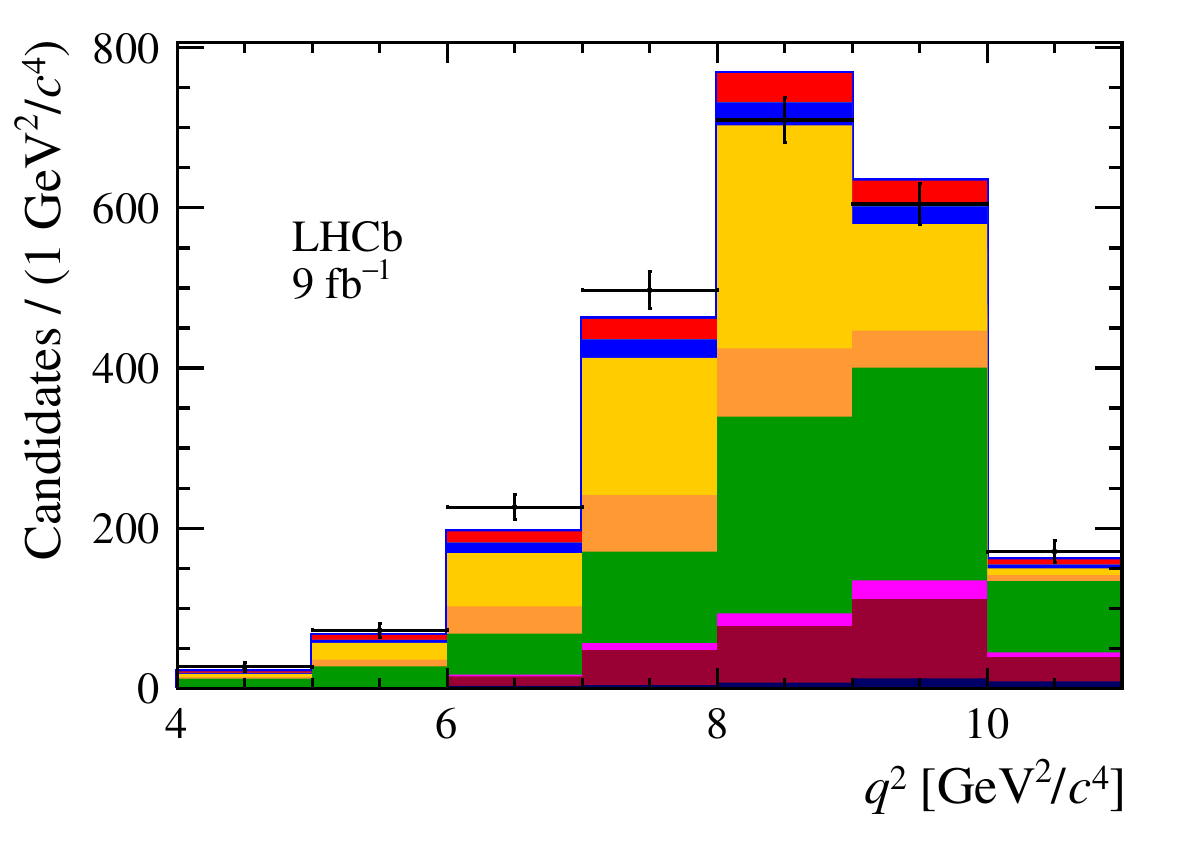}
\caption{ (Top) Distribution of the  mass difference between the \Ddoublez mass and the \Dstarp mass, (bottom left)  BDT-anti$D_s$ output distribution, and  (bottom right) distribution of the \taum\neutb invariant mass squared, \qsq,  for all candidates fulfilling the final \Ddouble selection. The fake \Ddoublez component is constructed with the WS data sample.}
\label{fig:fit3D_before_deltaM}
\end{figure}

As in Ref.~\cite{LHCb-PAPER-2022-052}, the determination of the signal yield is performed  using a three-dimensional binned maximum-likelihood fit to the output of the BDT-anti$D_s$ classifier,  the squared mass of the $\taum\neutb$ system, \qsq, and  the mass difference  \dm. This latter variable  is used to differentiate between \mbox{$\Bm\to\dsumz\taum\neutb$} and  \mbox{$\Bm\to\doneprimz\taum\neutb$} signals while the  BDT-anti$D_s$ classifier separates background contributions such as \mbox{$\Bm\to\Ddoublez\Dsm(X)$} decays from signal decays. Given the limited size of the data sample and to reduce the number of empty bins, the fit uses 8 bins for \dm, 5 for the BDT-anti$D_s$ output, and 3 for \qsq.  The \qsq variable is computed exactly as in Ref.~\cite{LHCb-PAPER-2022-052}. Taking into account the extra pion from the \Ddoublez decay  leads to the same signal yield within 2\%, taken into account in the systematic uncertainty assigned to the fitting procedure.
Simulation and control samples are used to determine the  templates that contribute to the final fit:  the \mbox{\Bm\to\Ddoublez\taum\neutb}, \mbox{$\Bm\to\Ddoublez\Dsm(X)$} and {\it prompt} \mbox{$B\to\Ddoublez\pim\pip\pim X$} simulated samples,  the WS control sample, together with two simulated samples of \mbox{$B\to\Dstarp\Dstarm (X)$} and \mbox{$B$\to$D^{*+}(DK)^-(X)$} decays. 
 Seven  \mbox{$\Bz\to\Dstarp\Dsm(X)$} templates  are used in total, of which six account for the  combinations of \Dsm and \Dssm mesons with each of the three \Ddoublez mesons, and one includes all \Dsssm contributions.
The signal templates are computed for each of the three \donez, \dtwoz and \doneprimz states.  Given the small mass difference between the \donez and the \dtwoz states, a common template is used for the sum of the two signal components. Their relative contribution is obtained from their predicted  production ratio and branching fractions to \Dstarp\pim final states~\cite{rmp}.   The relative contribution  of \doneprimz with respect to  \dsumz is constrained to the theoretical prediction of $0.4\pm0.5$~\cite{BLR,gubernari}.
The yields of five components (signal, \mbox{$\Bm\to\donez \Dssm$}, \mbox{$B\to\Dstarp\Dstarm (X)$}, \mbox{$B\to\Dstarp DK(X)$}, and the WS templates) are free to float in the fit, while  the remaining  yields are constrained to the corresponding estimated values. 
Figure~\ref{fig:fit3D_before_deltaM} shows the distributions of the three observables, together with the fit projections. The \chisq per degree of freedom is 0.89, with 115 degrees of freedom, demonstrating a good understanding of the various backgrounds. The   \mbox{$\Bm\to\dsumz\taum \neutb$} and \mbox{$\Bm\to\Ddoublez\taum \neutb$} decays yields are  found to be  $123\pm 23$ and   $220 \pm 34$, respectively. The quoted uncertainty is statistical, and includes the effects of the Gaussian constraints used in the fit.

The significance of the decay \decay{\Bm}{\Ddoublez\taum\neutb}, including a possible contribution from \doneprimz\taum\neutb, is determined to be 3.5\,$\sigma$, including the systematic uncertainties related to the signal yield determination discussed below. The significance level is  obtained comparing the results of a fit to the data allowing the contributions from \decay{\Bm}{\Ddoublez\Dsssm} and the relative rates of \donez and \dtwoz to vary freely, against those of a fit including only background.

The signal yield is compared to the normalisation channel yield to derive the branching fraction ratio \mbox{$\kappa_{\dsumz}$}, through the relation
\begin{equation}
 \label{eqr}
\kappa_{\dsumz}\equiv\frac{\BR(\Bm\to\dsumz\taum\neutb)}{\BR(\Bm\to\dsumz D_{s}^{(*)-})} =\frac{N_{\text{signal}}}{N_{\text{norm}}}\times\frac{\BR(\Dsm\to\pim\pim\pip)}{\BR(\taum\to\pim\pim\pip(X))\times\epsilon_{R}},
\end{equation}
where $N_{\text{signal}}$ is the signal yield, $N_{\text{norm}}$ denotes the yield of the normalisation  channel, and $\epsilon_{R}$ is the ratio of selection efficiencies between the two channels.

Various effects are studied to assess the systematic uncertainty  on the branching fraction ratio \mbox{$\kappa_{\dsumz}$}. In order to check the fit stability and potential bias, the same fit procedure is performed on selected candidates from the  simulated sample containing  inclusive $b$-hadron decays to final states containing  \mbox{$\Dstar\pim\pip\pim$}, and with a signal-to-background ratio similar to that observed in  data,  and with an equivalent luminosity 10 times larger. The  \dsumz signal yield is found to be compatible within $\pm1\sigma$ with the true signal yield in the simulated sample. Additional studies are performed with progressively decreasing signal yields. 
In all cases, the extracted signal yields are found to be in good agreement with the number of signal candidates present in the simulated samples. In an extreme scenario where no signal candidates are simulated, the fit returned a result of $35\pm80$, which is consistent with zero, corresponding to a potential yield in the data of 4$\pm$25 events. 
In the simulation, the three \mbox{$\Bm \to\Ddoublez\taum\neutb$} decays are generated using the Isgur--Scora--Grinstein--Wise (ISGW2) model~\cite{ISGW}. More recent models  better describe these semileptonic decays: Leibovich--Ligeti--Stewart--Wise (LLSW)  \cite{LLSW} and Bernlochner--Ligeti--Robinson (BLR) \cite{BLR}. Alternative templates are obtained for the main \mbox{$\Bm\to \donez \taum\neutb$} contribution by using the Hammer package~\cite{Hammer} to weight events according to these two models. The \qsq distributions for each model are shown in Fig.~\ref{fig:q_hammer}. The yields obtained with the LLSW and BLR models are found to be compatible  with the baseline result using the ISGW2 model, and the 3.7\% relative difference found is assigned as the systematic uncertainty due to the limited knowledge of the form-factor distributions. 
The fit assumes a fixed ratio between the \donez and \dtwoz signal contributions:  a systematic uncertainty of 4.4\% is assigned by varying their ratio from 0 to 1. The fit \chisq increases for large \dtwoz contributions,  a pure \dtwoz component is excluded at the level of $2.7 \sigma$, and a 50\% \dtwoz component is disfavoured at the $2 \sigma$ level.

The  systematic uncertainty associated with the finite size of the simulated  samples leads to an uncertainty of 4.3\% for the efficiency computation and 4.1\% for the template shapes determined using the  bootstrap method~\cite{efron}, as in Ref.~\cite{LHCb-PAPER-2022-052}.
The effect of changing fit observables, replacing the \qsq variable  by the \Dstar helicity angle~\cite{LHCb-PAPER-2023-020}, and changing the binning schemes with half or twice the number of  bins, results in a systematic uncertainty of 5.0\%, corresponding to half the maximum deviation from the baseline result.
The systematic uncertainty associated with the  \mbox{$\Bm\to\Ddoublez(DK)^-$}  background contamination is estimated to be 3.6\%, under the assumption that half of the estimated yield consists of signal candidates. 

This analysis requires the  reconstruction of an extra track, compared to the ${\cal R}(\Dstar)$ analysis, which has studied the efficiency in great detail~\cite{LHCb-PAPER-2022-052}. The uncertainty associated to the extra track reconstruction efficiency cancels to a large extent since the normalisation channel also requires an extra track reconstruction.  
The \mbox{$\Bm\to\Dstarp\pim\pim\pim\pip$} control sample is used to compare the selection efficiency in data and simulation. For each selection criterion the relative efficiency is measured, using data around the known \Bm mass~\cite{PDG2024}. A good agreement is found, except for the efficiency of the vertex detachment requirement, which deviates by 10\% due to differences in the vertex resolution between  data and simulation at 13 TeV. After correcting for such discrepancy, a systematic uncertainty of 2\% is assigned due to the limited knowledge of the analysis selection efficiency, in line with what has been found in the $ {\cal R(}\Dstar)$ analysis~\cite{LHCb-PAPER-2022-052}.

The background suppression due to  the detachment requirement must be correctly described in simulation, so that the  contamination of prompt background is correctly accounted in the fit. A comparison is made between  the suppression factor in data and in simulation of the prompt \mbox{$\Bm\to\Dstarp\pim\pim\pim\pip$} peak when requiring 2, 3 or  4$\,\sigma$ separation between the \Bm and \taum vertices. An agreement  within 4\% is observed and assigned as a systematic uncertainty.
A key feature of this analysis is the modelling of the fake \Ddouble background with data using the WS sample. The discrepancy between the observed  background under the \dm peak and its estimation is less than 5\% in the full \dm range (\cf Fig.~\ref{fig:deltaM_WS_check}).  Weighting the WS background mass distribution within such limit leads to a signal yield change of 2\%, which is assigned as the corresponding systematic uncertainty.
Table~\ref{tab:syssum} summarises the systematic uncertainties for this measurement. The total uncertainty is the sum in quadrature of the individual components. 
\begin{table}[]
\centering
\caption{Summary of relative systematic uncertainties on $\kappa_{\dsumz}$ related to the various identified sources.}
\label{tab:syssum}
 \begin{tabular}{l r}
    \hline
Source&Systematic uncertainty [\%]\\
\hline
Form factors&3.7\\
\dtwoz fraction&4.4\\
Finite size of the simulated samples - efficiency&4.3\\
Finite size of the simulated samples - templates&4.1\\

Variables and binning choices&5.0\\
\mbox{$\Bm\to\Ddoublez(DK)^-$} background&3.6\\
Selection and analysis&2.0\\
Detachment requirement&4.0\\
WS background description&2.0\\
   \hline
   Total& 11.4\\
   \hline
 \end{tabular}
\end{table}

In order to improve the statistical precision of the normalisation channel yield, the \mbox{$\Bm\to\Dstarp\Dsm\pim$} and  \mbox{$\Bm\to\Dstarp\Dssm(\to\Dsm\gamma)\pim$} contributions  are summed. The respective yields, obtained after a sideband subtraction around the \Dsm mass peak, are $116\pm11$ and $177\pm14$.   
The ratio of the corresponding branching fractions has been measured by the \lhcb collaboration to be $1.31 \pm 0.16$~\cite{LHCb-PAPER-2024-001}. The efficiency ratio $\epsilon_{R}$ between the signal and normalisation channels, is expected to be  far away from unity because of the presence of the two neutrinos in the signal channel. It is measured to be $\epsilon_{R}=0.38\pm0.02$ in the inclusive simulated sample,  taking  into account the slightly different efficiencies of the \Dsm and \Dssm channels. 
Combining these informations in Eq.~\ref{eqr} gives 
$$\frac{\BR(\Bm\to\dsumz\taum\neutb)}{\BR(\Bm\to\dsumz D_{s}^{(*)-})}=0.19\pm0.04\stat\pm0.02\syst. \nonumber$$

Using the branching fraction of the normalisation channel from Ref.~\cite{LHCb-PAPER-2024-001}, \mbox{$\BR(\Bm \to\dsumz D_s^{(*)})=(0.27 \pm 0.05)\%$}, where \donez and \dtwoz mesons decay to the \Dstarp\pim final state, it follows that
\begin{equation*}
\BR(\Bm\to\dsumz\taum\neutb)\times\BR(\dsumz\to\Dstarp\pim) = (0.051\pm0.013\stat\pm 0.006\syst \pm 0.009(\rm{ext}) )\%,
\end{equation*}
where the last uncertainty accounts for  the normalisation channel branching fraction uncertainty.
The branching fraction $\BR(\Bm\to\Ddoublez\taum\neutb)\times\BR(\Ddoublez\to\Dstarp\pim)= (0.085\pm0.026\stat\pm 0.009\syst \pm 0.016(\rm{ext}) )$\% corresponds to the average total \Ddoublez yield in the two extreme cases where the fraction of the \doneprimz component in the \Ddoublez\taum\neutb signal is assumed to be 0 or 100\%,  and the first uncertainty covers the variation of the total \Ddoublez yield between these two cases. The branching fraction $\BR(\Bm\to\dsumz\taum\neutb)\times\BR(\dsumz\to\Dstarp\pim)$ is found to be  less than 0.094\% at 90\% C.L. in the first case. In the very unlikely case of a pure \doneprimz\taum\neutb component, the result is $\BR(\Bm\to\doneprimz\taum\neutb)\times\BR(\doneprimz\to\Dstarp\pim)=(0.111\pm0.017\stat\pm 0.010\syst \pm 0.019(\rm{ext}))\%$.
Given that the branching fractions for the  muonic \Bm-decay channels involving the  \donez and \dtwoz states are $(0.30\pm0.02)\%$ and $(0.10\pm0.02)\%$~\cite{PDG2024}, the following ratio is determined 
 \begin{equation}  
 \label{eq4}
{\cal R}(\dsumz)=0.13\pm0.03\stat\pm0.01\syst\pm0.02\,(\rm{ext}).\nonumber
 \end{equation}
The measured ratio is  compatible with the SM prediction of $0.09\pm0.02$~\cite{BLR, gubernari,roudeau} and with assumptions made in the ${\cal R}(D^{(*)})$ measurements~\cite{Lees_2012,belle_collaboration_measurement_2015,belle-2016, the_belle_collaboration_measurement_2018,LHCb-PAPER-2017-017,LHCb-PAPER-2017-027, LHCb-PAPER-2022-052, LHCb-PAPER-2022-039, LHCb-PAPER-2024-007,belle-II} on the ${\cal R}(\Ddouble)$ value. 
This implies that the observed deviation of ${\cal R}(D^{(*)})$ from the SM prediction is not likely to be due to a global underestimation of  \mbox{$B\to\Ddouble\taum\neutb$} decays. 

Regarding the ${\cal R}(\Dstar)$ measurement by LHCb collaboration based on hadronic \taum decays~\cite{LHCb-PAPER-2022-052}, and assuming a total of 9000 \mbox{$\Bzb\to\Dstarp\taum\neutb$} decays in the full 9\invfb data sample,  the   expected value of the feed-down fraction  of \mbox{$B\to\Ddouble\taum\neutb$} candidates in the fitted sample is  estimated to be $(8.9\pm 2.1)\%$, leading to an  upper limit of  11.1\% (13.1\%)  at the 90\% (95\%) confidence level. Taking into account the extra $(0.7\pm0.1)\%$  contribution arising from \mbox{$\Bsb\to\Dsssp\taum\neutb$} decays and the uncertainty regarding the total signal yield, the total feed-down contribution is larger than the 3.5\% rate used in Ref.\cite{LHCb-PAPER-2022-052}, but compatible within $2.6\sigma$. The corresponding shift in  $ {\cal R(}\Dstar)$ quoted there is well within one standard deviation.

In summary, evidence for the  \mbox{$\Bm\to\Ddoublez\taum\neutb$} channel is obtained with a significance  of $3.5\,\sigma$. The branching fractions 
$\BR(\Bm\to\dsumz\taum\neutb)\times\BR(\dsumz\to\Dstarp\pim)$ and the ratio  ${\cal R(}\dsumz)$ are measured for the first time. The latter is  found to be in good agreement with the SM prediction and consistent with the assumptions used in the various ${\cal R}(D^{(*)})$ measurements.
 \section*{End matters}
\begin{figure}[b]
\begin{center}
\includegraphics[width=0.8\linewidth]{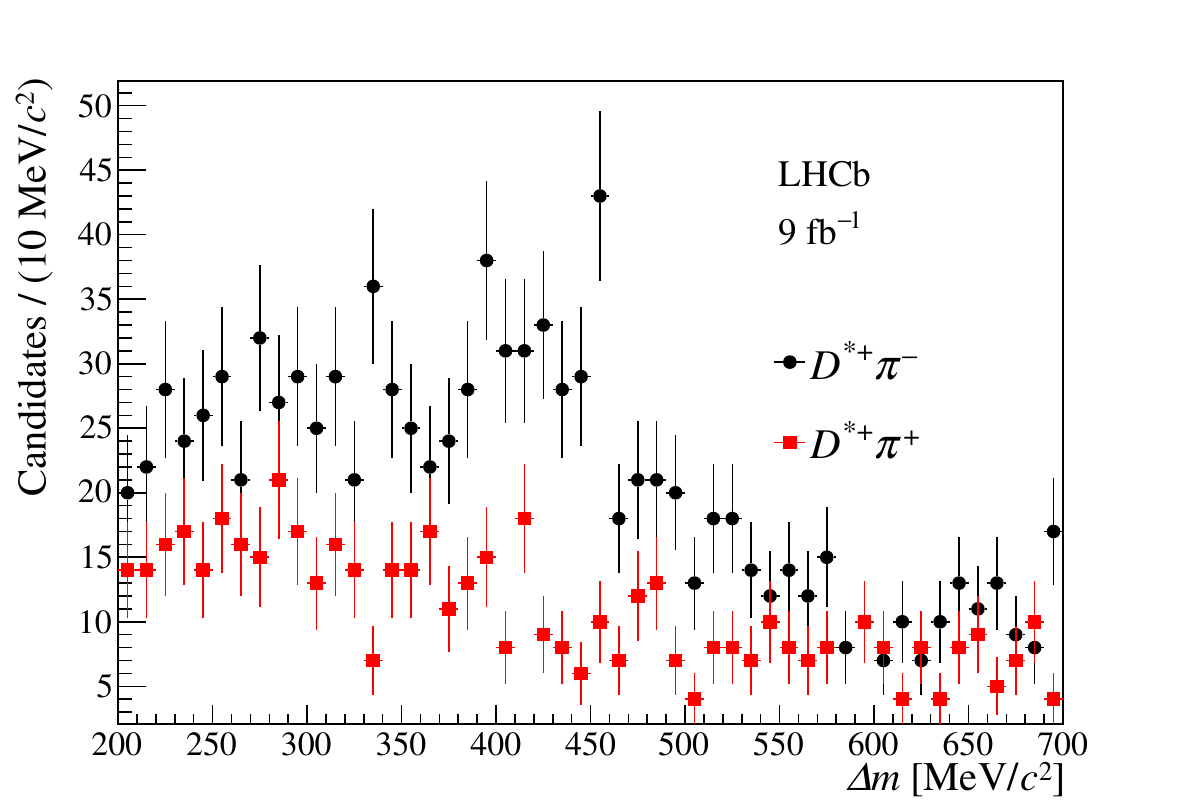}
\caption{Distribution of the \dm  mass difference for \mbox{$\Bm\to\Dstarp\pim\Dzb (\Km)$} candidates, with  \Dzb  reconstructed in the \Kp\pim\pim\pip final state.~Here the \Km meson is not reconstructed. The black (red) points correspond to the RS (WS) sample.}
\label{fig:dz_deltam}
\end{center}
\end{figure}

In order  to ascertain whether genuine \mbox{$B\to\Ddoublez\Dzb X$} events are present when associated to fully reconstructed \Dzb candidates, a sample of events combining fully reconstructed \Dzb and \Ddoublez candidates is selected. Candidate \Dzb mesons are reconstructed in the \Kp\pim\pim\pip mode, requiring the presence of an extra kaon originating from the three pion vertex. Figure~\ref{fig:dz_deltam} shows the \dm distribution for the \Ddoublez candidates reconstructed together with a \Dzb meson. In the distribution of the mass difference between the \donez and \Dstarp mesons, only  a small hint of \Ddoublez candidates is observed around 410 \mevcc.  This demonstrates that the vast majority of these events contains fake \Ddoublez candidates.

\begin{figure}[!tb]
\begin{center}
\includegraphics[width=0.8\linewidth]{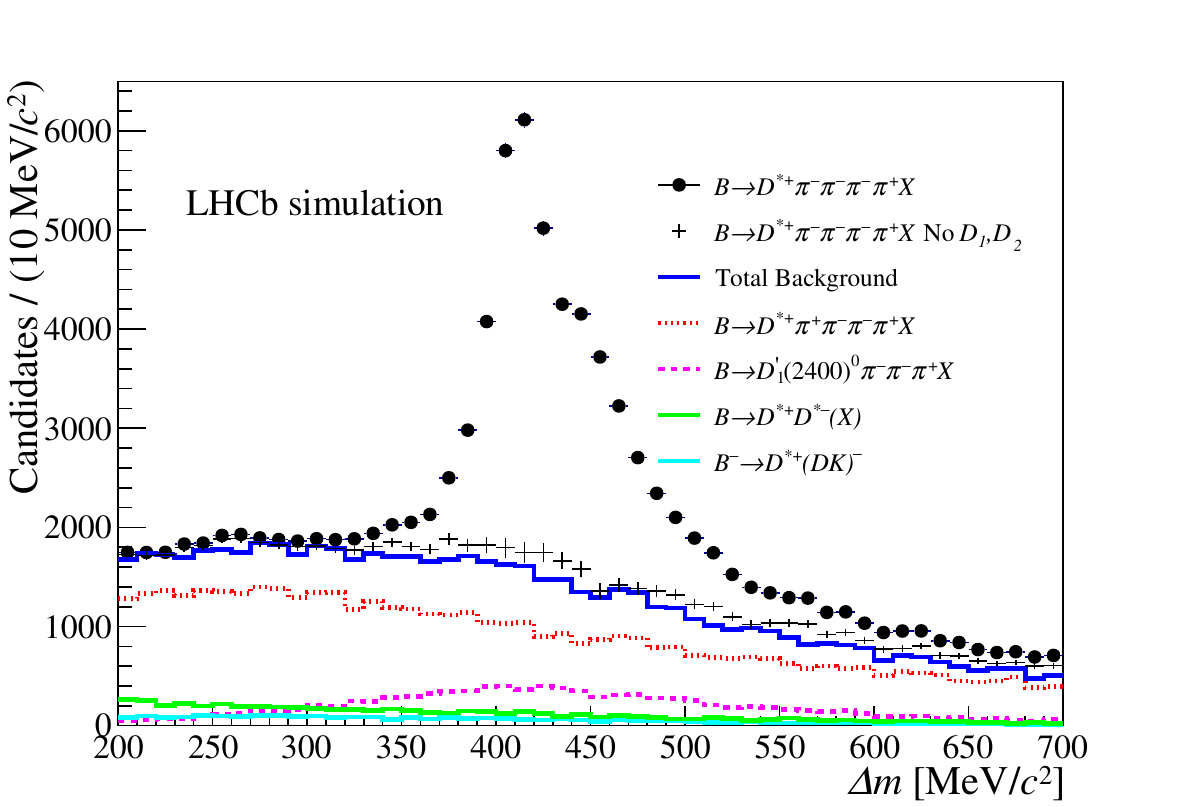}
\caption{Distribution of the \dm mass difference for all selected \Ddoublez candidates in the inclusive simulated sample (black circles) compared to the WS sample extracted from the same dataset. The background under the \dm peak in RS sample (black crosses) corresponds to RS simulated events not containing a \donez or \dtwoz meson. A good agreement is found between this component and the  (dark blue) total background estimate which is the sum of the WS candidates plus RS candidates containing either  a \doneprimz meson, a \Dstarp\Dstarm pair, or  three-body \mbox{$\Dstarp(DK)^{-}$} in the final state. }
\label{fig:deltaM_WS_check}
\end{center}
\end{figure}

In order to check that the fake \Ddoublez background can be well described by the \Ddoublez WS sample, a study is performed using a simulated sample of inclusive $b$-hadron decays to final states containing  a \Dstarp meson and three additional charged pions. The mass distribution of the background under the \Ddoublez peak in the RS sample is compared to that of the WS candidates in the same simulated sample. As can be seen in Fig.~\ref{fig:deltaM_WS_check}, a good agreement is found, validating the procedure.

\begin{figure}[]
\begin{center}
\includegraphics[width=0.8\linewidth]{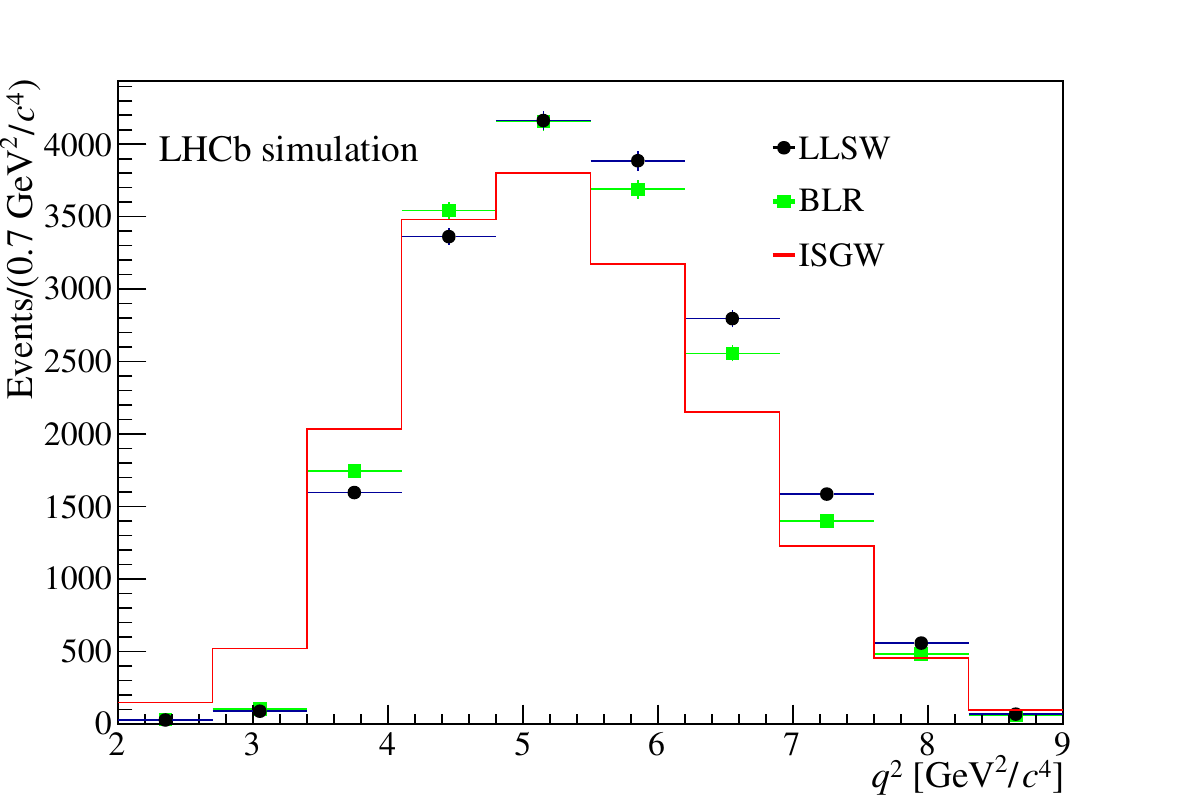}
\caption{Distribution of the \qsq  observable for simulated \mbox{$\Bm\to\donez\taum\neutb$} decays corresponding to the BLR~\cite{BLR}, LLSW~\cite{LLSW} and ISGW2~\cite{ISGW} form-factor models.}
\label{fig:q_hammer}
\end{center}
\end{figure}

Three different theoretical models (BLR~\cite{BLR}, LLSW~\cite{LLSW}, and ISGW2~\cite{ISGW}) can be used to describe the semileptonic  $B$\to\Ddouble\tauon\neu decays. Figure~\ref{fig:q_hammer} displays the \qsq distribution for each model, showing that the BLR and LLSW models are rather similar, while the ISGW2 model predicts a \qsq distribution shifted toward lower values.

\clearpage
\input{acknowledgements}

\clearpage
\addcontentsline{toc}{section}{References}
\bibliographystyle{LHCb}
\bibliography{main,standard,LHCb-PAPER,LHCb-CONF,LHCb-DP,LHCb-TDR}
\clearpage
\newpage
\input{Authorship_LHCb-PAPER-2024-037}
\end{document}

%% file: preamble.tex
\usepackage[top=1in, bottom=1.25in, left=1in, right=1in]{geometry}

\usepackage{microtype}
\usepackage{lineno}  
\usepackage{xspace} 
\usepackage{caption} 

\usepackage{graphicx}  
\usepackage{color}
\usepackage{colortbl}
\graphicspath{{./figs/}} 

\usepackage{amsmath} 
\usepackage{amssymb}
\usepackage{amsfonts}
\usepackage{upgreek} 

\newcommand*\patchAmsMathEnvironmentForLineno[1]{%
\expandafter\let\csname old#1\expandafter\endcsname\csname #1\endcsname
\expandafter\let\csname oldend#1\expandafter\endcsname\csname
end#1\endcsname
 \renewenvironment{#1}%
   {\linenomath\csname old#1\endcsname}%
   {\csname oldend#1\endcsname\endlinenomath}%
}
\newcommand*\patchBothAmsMathEnvironmentsForLineno[1]{%
  \patchAmsMathEnvironmentForLineno{#1}%
  \patchAmsMathEnvironmentForLineno{#1*}%
}
\AtBeginDocument{%
\patchBothAmsMathEnvironmentsForLineno{equation}%
\patchBothAmsMathEnvironmentsForLineno{align}%
\patchBothAmsMathEnvironmentsForLineno{flalign}%
\patchBothAmsMathEnvironmentsForLineno{alignat}%
\patchBothAmsMathEnvironmentsForLineno{gather}%
\patchBothAmsMathEnvironmentsForLineno{multline}%
\patchBothAmsMathEnvironmentsForLineno{eqnarray}%
}

\usepackage{hyperxmp}

\usepackage[pdftex,
            pdfauthor={\paperauthors},
            pdftitle={\paperasciititle},
            pdfkeywords={\paperkeywords},
            pdfcopyright={Copyright (C) \papercopyright},
            pdflicenseurl={\paperlicenceurl}]{hyperref}

\usepackage[colorinlistoftodos,textsize=scriptsize]{todonotes}

\usepackage[bottom,flushmargin,hang,multiple]{footmisc}

\usepackage[all]{hypcap} 

\input{lhcb-symbols-def} 
\usepackage{float}
\usepackage{cite} 
\usepackage{mciteplus}

%% file: lhcb-symbols-def.tex
\usepackage{xspace} 
\usepackage{upgreek}

\def\lhcb   {\mbox{LHCb}\xspace}

\def\babar  {\mbox{BaBar}\xspace}

\def\besiii {\mbox{BESIII}\xspace}

\def\MagUp {\mbox{\em Mag\kern -0.05em Up}\xspace}

\ifthenelse{\boolean{uprightparticles}}%
{

 \def\Pmu         {\ensuremath{\upmu}\xspace}                 
 \def\Pnu         {\ensuremath{\upnu}\xspace}                 
                  
 \def\Ppi         {\ensuremath{\uppi}\xspace}

 \def\Ptau        {\ensuremath{\uptau}\xspace}

 \def\PDelta      {\ensuremath{\Delta}\xspace}                 
 \def\PXi         {\ensuremath{\Xi}\xspace}                 
 \def\PLambda     {\ensuremath{\Lambda}\xspace}                 
 \def\PSigma      {\ensuremath{\Sigma}\xspace}                 
 \def\POmega      {\ensuremath{\Omega}\xspace}                 
 \def\PUpsilon    {\ensuremath{\Upsilon}\xspace}

 \def\PB      {\ensuremath{\mathrm{B}}\xspace}                 
                  
 \def\PD      {\ensuremath{\mathrm{D}}\xspace}

 \def\PK      {\ensuremath{\mathrm{K}}\xspace}

 \def\Pb      {\ensuremath{\mathrm{b}}\xspace}                 
 \def\Pc      {\ensuremath{\mathrm{c}}\xspace}                 
 \def\Pd      {\ensuremath{\mathrm{d}}\xspace}

 \def\Pi      {\ensuremath{\mathrm{i}}\xspace}

 \def\Pp      {\ensuremath{\mathrm{p}}\xspace}

 \def\Ps      {\ensuremath{\mathrm{s}}\xspace}                 
                  
 \def\Pu      {\ensuremath{\mathrm{u}}\xspace}

 \def\thebaroffset{0.0em}
}
{

 \def\Pmu         {\ensuremath{\mu}\xspace}                 
 \def\Pnu         {\ensuremath{\nu}\xspace}                 
                  
 \def\Ppi         {\ensuremath{\pi}\xspace}

 \def\Ptau        {\ensuremath{\tau}\xspace}

 \mathchardef\PDelta="7101
 \mathchardef\PXi="7104
 \mathchardef\PLambda="7103
 \mathchardef\PSigma="7106
 \mathchardef\POmega="710A
 \mathchardef\PUpsilon="7107
                  
 \def\PB      {\ensuremath{B}\xspace}                 
                  
 \def\PD      {\ensuremath{D}\xspace}

 \def\PK      {\ensuremath{K}\xspace}

 \def\Pb      {\ensuremath{b}\xspace}                 
 \def\Pc      {\ensuremath{c}\xspace}                 
 \def\Pd      {\ensuremath{d}\xspace}

 \def\Pi      {\ensuremath{i}\xspace}

 \def\Pp      {\ensuremath{p}\xspace}

 \def\Ps      {\ensuremath{s}\xspace}                 
                  
 \def\Pu      {\ensuremath{u}\xspace}

 \def\thebaroffset{0.18em}
}
\newcommand{\offsetoverline}[2][\thebaroffset]{\kern #1\overline{\kern -#1 #2}}%

\makeatletter
\ifcase \@ptsize \relax
  \newcommand{\miniscule}{\@setfontsize\miniscule{4}{5}}
\or
  \newcommand{\miniscule}{\@setfontsize\miniscule{5}{6}}
\or
  \newcommand{\miniscule}{\@setfontsize\miniscule{5}{6}}
\fi
\makeatother

\DeclareRobustCommand{\optbar}[1]{\shortstack{{\miniscule (\rule[.5ex]{1.25em}{.18mm})}
  \\ [-.7ex] $#1$}}

\def\mun        {{\ensuremath{\Pmu^-}}\xspace} 

\def\tauon      {{\ensuremath{\Ptau}}\xspace}

\def\taum       {{\ensuremath{\Ptau^-}}\xspace}

\def\neu        {{\ensuremath{\Pnu}}\xspace}
\def\neub       {{\ensuremath{\overline{\Pnu}}}\xspace}

\def\neumb      {{\ensuremath{\neub_\mu}}\xspace}
\def\neut       {{\ensuremath{\neu_\tau}}\xspace}
\def\neutb      {{\ensuremath{\neub_\tau}}\xspace}

\def\uquark    {{\ensuremath{\Pu}}\xspace}
\def\uquarkbar {{\ensuremath{\overline \uquark}}\xspace}

\def\dquark    {{\ensuremath{\Pd}}\xspace}
\def\dquarkbar {{\ensuremath{\overline \dquark}}\xspace}

\def\squark    {{\ensuremath{\Ps}}\xspace}

\def\cquark    {{\ensuremath{\Pc}}\xspace}

\def\bquark    {{\ensuremath{\Pb}}\xspace}

\def\pion   {{\ensuremath{\Ppi}}\xspace}
\def\piz    {{\ensuremath{\pion^0}}\xspace}
\def\pip    {{\ensuremath{\pion^+}}\xspace}
\def\pim    {{\ensuremath{\pion^-}}\xspace}

\def\kaon    {{\ensuremath{\PK}}\xspace}

\def\KorKbar {\kern \thebaroffset\optbar{\kern -\thebaroffset \PK}{}\xspace}

\def\Kp      {{\ensuremath{\kaon^+}}\xspace}
\def\Km      {{\ensuremath{\kaon^-}}\xspace}

\def\Dbar    {{\ensuremath{\offsetoverline{\PD}}}\xspace}
\def\D       {{\ensuremath{\PD}}\xspace}

\def\DorDbar {\kern \thebaroffset\optbar{\kern -\thebaroffset \PD}\xspace}
\def\Dz      {{\ensuremath{\D^0}}\xspace}
\def\Dzb     {{\ensuremath{\Dbar{}^0}}\xspace}
\def\Dp      {{\ensuremath{\D^+}}\xspace}
\def\Dm      {{\ensuremath{\D^-}}\xspace}

\def\DpDm    {\ensuremath{\Dp {\kern -0.16em \Dm}}\xspace}
\def\Dstar   {{\ensuremath{\D^*}}\xspace}

\def\Dstarp  {{\ensuremath{\D^{*+}}}\xspace}
\def\Dstarm  {{\ensuremath{\D^{*-}}}\xspace}

\def\Ds      {{\ensuremath{\D^+_\squark}}\xspace}

\def\Dsm     {{\ensuremath{\D^-_\squark}}\xspace}

\def\Dsssm     {{\ensuremath{\D^{**-}_\squark}}\xspace}
\def\Dsssp     {{\ensuremath{\D^{**+}_\squark}}\xspace}

\def\Dssm    {{\ensuremath{\D^{*-}_\squark}}\xspace}

\def\Ddouble {{\ensuremath{\D^{**}}}\xspace}
\def\Ddoublez {{\ensuremath{\D^{**0}}}\xspace}

\def\ddzero {{\ensuremath{\D_{0}^{*}(2300)}}\xspace}

\def\done {{\ensuremath{\D_{1}(2420)}}\xspace}
\def\donez {{\ensuremath{\D_{1}(2420)^{0}}}\xspace}

\def\doneprim {{\ensuremath{\D_{1}^{'}(2400)}}\xspace}
\def\doneprimz{{\ensuremath{\D_{1}^{'}(2400)^{0}}}\xspace}

\def\dtwo {{\ensuremath{\D_{2}^{*}(2460)}}\xspace}
\def\dtwoz {{\ensuremath{\D_{2}^{*}(2460)^{0}}}\xspace}

\def\dsumz{{\ensuremath{\D_{1,2}^{**0}}}\xspace}

\def\B       {{\ensuremath{\PB}}\xspace}
\def\Bbar    {{\ensuremath{\offsetoverline{\PB}}}\xspace}

\def\BorBbar {\kern \thebaroffset\optbar{\kern -\thebaroffset \PB}\xspace}
\def\Bz      {{\ensuremath{\B^0}}\xspace}
\def\Bzb     {{\ensuremath{\Bbar{}^0}}\xspace}
\def\Bd      {{\ensuremath{\B^0}}\xspace}

\def\BdorBdbar {\kern \thebaroffset\optbar{\kern -\thebaroffset \Bd}\xspace}

\def\Bub     {{\ensuremath{\B^-}}\xspace}

\def\Bm      {{\ensuremath{\Bub}}\xspace}

\def\Bs      {{\ensuremath{\B^0_\squark}}\xspace}
\def\Bsb     {{\ensuremath{\Bbar{}^0_\squark}}\xspace}
\def\BsorBsbar {\kern \thebaroffset\optbar{\kern -\thebaroffset \Bs}\xspace}

\def\Y#1S{\ensuremath{\PUpsilon{(#1S)}}\xspace}

\def\proton      {{\ensuremath{\Pp}}\xspace}

\def\LorLbar     {\kern \thebaroffset\optbar{\kern -\thebaroffset \PLambda}\xspace}

\def\BF         {{\ensuremath{\mathcal{B}}}\xspace}
\def\BR         {\BF}

\newcommand{\decay}[2]{\ensuremath{#1\!\to #2}\xspace} 

\def\to                 {\ensuremath{\rightarrow}\xspace}

\def\qsq       {{\ensuremath{q^2}}\xspace}

\newcommand{\dm}{{\ensuremath{\Delta m}}\xspace}

\def\AT#1     {\ensuremath{A_{\mathrm{T}}^{#1}}\xspace}           

\def\C#1      {\ensuremath{\mathcal{C}_{#1}}\xspace}                       
\def\Cp#1     {\ensuremath{\mathcal{C}_{#1}^{'}}\xspace}                    
\def\Ceff#1   {\ensuremath{\mathcal{C}_{#1}^{\mathrm{(eff)}}}\xspace}        
\def\Cpeff#1  {\ensuremath{\mathcal{C}_{#1}^{'\mathrm{(eff)}}}\xspace}       
\def\Ope#1    {\ensuremath{\mathcal{O}_{#1}}\xspace}                       
\def\Opep#1   {\ensuremath{\mathcal{O}_{#1}^{'}}\xspace}                    

\newcommand{\nospaceunit}[1]{\ensuremath{\text{#1}}}       
\newcommand{\aunit}[1]{\ensuremath{\text{\,#1}}}       

\newcommand{\tev}{\aunit{Te\kern -0.1em V}\xspace}
\newcommand{\gev}{\aunit{Ge\kern -0.1em V}\xspace}
\newcommand{\mev}{\aunit{Me\kern -0.1em V}\xspace}
\newcommand{\kev}{\aunit{ke\kern -0.1em V}\xspace}
\newcommand{\ev}{\aunit{e\kern -0.1em V}\xspace}
 
\newcommand{\mevc}{\ensuremath{\aunit{Me\kern -0.1em V\!/}c}\xspace}
\newcommand{\gevc}{\ensuremath{\aunit{Ge\kern -0.1em V\!/}c}\xspace}
\newcommand{\mevcc}{\ensuremath{\aunit{Me\kern -0.1em V\!/}c^2}\xspace}
\newcommand{\gevcc}{\ensuremath{\aunit{Ge\kern -0.1em V\!/}c^2}\xspace}

\def\mum  {\ensuremath{\,\upmu\nospaceunit{m}}\xspace}

\def\fb   {\ensuremath{\aunit{fb}}\xspace}
\def\invfb   {\ensuremath{\fb^{-1}}\xspace}

\newcommand{\stat}{\aunit{(stat)}\xspace}
\newcommand{\syst}{\aunit{(syst)}\xspace}

\newcommand{\chisq}{\ensuremath{\chi^2}\xspace}

\def\gsim{{~\raise.15em\hbox{$>$}\kern-.85em
          \lower.35em\hbox{$\sim$}~}\xspace}
\def\lsim{{~\raise.15em\hbox{$<$}\kern-.85em
          \lower.35em\hbox{$\sim$}~}\xspace}

\def\pt         {\ensuremath{p_{\mathrm{T}}}\xspace}

\def\evtgen     {\mbox{\textsc{EvtGen}}\xspace}

\def\geant      {\mbox{\textsc{Geant4}}\xspace}

\def\photos     {\mbox{\textsc{Photos}}\xspace}

\def\pythia     {\mbox{\textsc{Pythia}}\xspace}

\def\tell1  {TELL1\xspace}
\def\ukl1   {UKL1\xspace}

\newcommand{\cf}{\mbox{\itshape cf.}\xspace}



%% file: acknowledgements.tex
\section*{Acknowledgements}
%
%
\noindent We express our gratitude to our colleagues in the CERN
accelerator departments for the excellent performance of the LHC. We
thank the technical and administrative staff at the LHCb
institutes.
We acknowledge support from CERN and from the national agencies:
ARC and ARDC (Australia);
CAPES, CNPq, FAPERJ and FINEP (Brazil); 
MOST and NSFC (China); 
CNRS/IN2P3 (France); 
BMBF, DFG and MPG (Germany); 
INFN (Italy); 
NWO (Netherlands); 
MNiSW and NCN (Poland); 
MCID/IFA (Romania); 
MICIU and AEI (Spain);
SNSF and SER (Switzerland); 
NASU (Ukraine); 
STFC (United Kingdom); 
DOE NP and NSF (USA).
We acknowledge the computing resources that are provided by CERN, IN2P3
(France), KIT and DESY (Germany), INFN (Italy), SURF (Netherlands),
PIC (Spain), GridPP (United Kingdom), 
CSCS (Switzerland), IFIN-HH (Romania), CBPF (Brazil),
and Polish WLCG (Poland).
We are indebted to the communities behind the multiple open-source
software packages on which we depend.
Individual groups or members have received support from
Key Research Program of Frontier Sciences of CAS, CAS PIFI, CAS CCEPP, 
Fundamental Research Funds for the Central Universities, 
and Sci. \& Tech. Program of Guangzhou (China);
Minciencias (Colombia);
EPLANET, Marie Sk\l{}odowska-Curie Actions, ERC and NextGenerationEU (European Union);
A*MIDEX, ANR, IPhU and Labex P2IO, and R\'{e}gion Auvergne-Rh\^{o}ne-Alpes (France);
AvH Foundation (Germany);
ICSC (Italy); 
Severo Ochoa and Mar\'ia de Maeztu Units of Excellence, GVA, XuntaGal, GENCAT, InTalent-Inditex and Prog. ~Atracci\'on Talento CM (Spain);
SRC (Sweden);
the Leverhulme Trust, the Royal Society
 and UKRI (United Kingdom).

%% file: Authorship_LHCb-PAPER-2024-037.tex
\centerline
{\large\bf LHCb collaboration}
\begin
{flushleft}
\small
R.~Aaij$^{38}$\lhcborcid{0000-0003-0533-1952},
A.S.W.~Abdelmotteleb$^{57}$\lhcborcid{0000-0001-7905-0542},
C.~Abellan~Beteta$^{51}$,
F.~Abudin{\'e}n$^{57}$\lhcborcid{0000-0002-6737-3528},
T.~Ackernley$^{61}$\lhcborcid{0000-0002-5951-3498},
A. A. ~Adefisoye$^{69}$\lhcborcid{0000-0003-2448-1550},
B.~Adeva$^{47}$\lhcborcid{0000-0001-9756-3712},
M.~Adinolfi$^{55}$\lhcborcid{0000-0002-1326-1264},
P.~Adlarson$^{82}$\lhcborcid{0000-0001-6280-3851},
C.~Agapopoulou$^{14}$\lhcborcid{0000-0002-2368-0147}
% ...

{\footnotesize \it

% ...
$^{14}$Universit{\'e} Paris-Saclay, CNRS/IN2P3, IJCLab, Orsay, France\\
% ...
$^{38}$Nikhef National Institute for Subatomic Physics, Amsterdam, Netherlands\\
% ...
$^{47}$Instituto Galego de F{\'\i}sica de Altas Enerx{\'\i}as (IGFAE), Universidade de Santiago de Compostela, Santiago de Compostela, Spain\\
% ...
$^{51}$Physik-Institut, Universit{\"a}t Z{\"u}rich, Z{\"u}rich, Switzerland\\
% ...
$^{55}$H.H. Wills Physics Laboratory, University of Bristol, Bristol, United Kingdom\\
% ...
$^{57}$Department of Physics, University of Warwick, Coventry, United Kingdom\\
% ...
$^{60}$School of Physics and Astronomy, University of Glasgow, Glasgow, United Kingdom\\
$^{61}$Oliver Lodge Laboratory, University of Liverpool, Liverpool, United Kingdom\\
% ...
$^{69}$Syracuse University, Syracuse, NY, United States\\
% ...
$^{82}$Department of Physics and Astronomy, Uppsala University, Uppsala, Sweden, associated to $^{60}$\\
% ...
}
\end{flushleft}